\documentstyle[epsfig]{preprint}    

\newcommand{\dslash}{\partial \hskip -0.5em /}
\newcommand{\vslash}{v \hskip -0.5em /}
\newcommand{\bD}{{\bf D}}

\newcommand{\fract}[2]{{\textstyle\frac{#1}{#2}}}
\newcommand{\lapeq}{\stackrel{\scriptscriptstyle\raisebox{-2.5mm}{$<$}}
{\scriptscriptstyle \raisebox{-0.8mm}{$\sim$}}}

\newcommand{\fpt}{F^{\prime2}}

\newcommand{\sFt}{{{\rm sin}^2}F}

\begin{document}
\title{Soliton Models for the Nucleon and Predictions for the 
Nucleon Spin Structure\footnote{Lectures presented at the Advanced
Study Institute {\it Symmetry and Spin} Prague, 2001. }}
\authori{Herbert Weigel\footnote{Heisenberg--Fellow}}
\addressi{Institute for Theoretical Physics, T\"ubingen University,\\
Auf der Morgenstelle 14, D--72076 T\"ubingen, Germany}
\authorii{}     
\addressii{}
\authoriii{}     
\headtitle{Soliton Models for the Nucleon \ldots}
\headauthor{Herbert Weigel}  
\specialhead{Herbert Weigel: Soliton Models for the Nucleon \ldots}
\daterec{} 
\maketitle

\begin{abstract}
In these lectures the three flavor soliton approach for baryons is 
reviewed. Effects of flavor symmetry breaking in the baryon 
wave--functions on axial current matrix elements are discussed. 
A bosonized chiral quark model is considered to outline the computation 
of spin dependent nucleon structure functions in the soliton picture.
\end{abstract}

\section{Introduction}     
Many nucleon properties cannot be computed from first principles even 
though the fundamental theory that describes the strong interaction 
processes of hadrons, Quantum Chromodynamics (QCD), is well established.
In QCD hadrons are composites of quarks and gluons whose interactions 
are described within the non--abelian gauge theory of color $SU(3)$. However, 
perturbative techniques are not applicable in the low--energy region. 
It is therefore mandatory to consider models which can be deduced 
or at least motivated from QCD. In this context it has been fruitful to 
observe that QCD contains a hidden expansion parameter, $N_C$, the 
number of colors. Generalizing from $SU(3)$ to $SU(N_C)$ and assuming 
the confinement phenomena, QCD becomes equivalent to a theory of weakly 
interacting mesons in the limit of large $N_C$~\cite{tH74}. That is, 
the coupling constants of the meson interactions scale like $1/N_C$ 
while baryon masses and radii scale like $N_C$ and $N_C^{0}$, 
respectively~\cite{Wi79}. Meson Lagrangians may possess localized solutions 
to the field equations with finite field energy, the so--called solitons. 
Soliton energies scale inversely with the meson coupling and their 
extensions approach constants as the coupling increases. It was thus 
conjectured that baryons emerge as solitons in the effective meson theory 
that is equivalent to QCD~\cite{Wi79}. Although this meson theory 
cannot be derived from QCD, low--energy meson phenomenology provides 
sufficient constraints to model this theory. Especially chiral
symmetry and its breaking in the vacuum are essential. This introduces
non--linear interactions for the pions, the (would--be) Goldstone
bosons of chiral symmetry, via the chiral field
\begin{equation}
U={\rm exp}\left(i\vec{\tau}\cdot\vec{\pi}/f\right)\, .
\label{chifield}
\end{equation}
Effective Lagrangians are constructed from $U$ that are 
invariant under global chiral transformations $U\to L U R^\dagger$. As
$U^\dagger U=1$ at least two derivatives are required
\begin{equation}
{\cal L}_0=\frac{f^2}{4}{\rm tr}
\left(\partial_\mu U \partial^\mu U^\dagger\right)\, .
\label{lag0}
\end{equation}
Extracting the axial current 
$A_\mu^a=f\pi^a+{\cal O}(\vec{\pi}\hspace{0.2mm}^3)$
from ${\cal L}_0$ provides the electroweak coupling and determines 
the pion decay constant $f=f_\pi=93{\rm MeV}$ from $\pi\to\mu\nu$.
Having established such a chiral model, a finite energy soliton 
solution must be obtained and quantized to describe baryon states.
I will outline this approach in section 2. In section 3 I will consider 
three flavor extensions thereof with special emphasis on the role of 
flavor symmetry breaking~\cite{We96}. I will employ these methods to 
compute axial current matrix elements of baryons in section 4. These
matrix elements are essential ingredients for the description of the
nucleon spin structure~\cite{Ja96} as they reflect its various quark 
flavor contributions~\cite{El95} and they parameterize hyperon
beta--decay. The effects of flavor symmetry breaking will be essential 
to discuss the strange quark contribution. In section~5 I will discuss a 
different approach by starting from a simple model for the quark flavor 
dynamics. Using standard bosonization techniques~\cite{Eb86} this model 
is rewritten as a (non--local) meson field theory which can be shown to 
support chiral solitons~\cite{Al96}. As I will describe in sections~6
and~7, the quark degrees of freedom can be traced through the 
bosonization procedure for the amplitude of virtual Compton scattering. 
This paves the way to compute nucleon structure functions from the 
chiral quark soliton. Finally section 8 contains some concluding remarks.

\section{Baryons as Chiral Solitons}
Scaling considerations show that the model~(\ref{lag0}) does 
not contain stable soliton solutions. Therefore Skyrme added a 
stabilizing term~\cite{Sk61}
\begin{equation}
{\cal L}=\frac{f_\pi^2}{4}{\rm tr} 
\left[\partial_\mu U \partial^\mu U^\dagger\right]
+\frac{1}{32e^2}{\rm tr}\left(
\left[U^\dagger\partial_\mu U,U^\dagger\partial_\nu U\right]
\left[U^\dagger\partial^\mu U,U^\dagger\partial^\nu U\right]\right)\, ,
\label{lagsk}
\end{equation}
that is of fourth order in the derivatives but only quadratic in the 
time derivatives. There are other approaches to extend ${\cal L}_0$ 
such that stable soliton solutions exist, as {\it e.g.} including 
vector mesons~\cite{Ja88,Ka84}. Although such extensions appear 
physically more motivated, I will stick to the Skyrme model for 
pedagogical reasons when explaining the soliton picture for baryons.

The soliton solution to (\ref{lagsk}) assumes the famous hedgehog
shape
\begin{equation}
U_{\rm H}\left(\vec{r}\,\right)=
{\rm exp}\left(i\vec{\tau}\cdot\hat{r}F(r)\right)\,.
\label{hedgehog}
\end{equation}
The equations of motion become an ordinary second order differential
equation for the chiral angle $F(r)$ that are obtained by
extremizing the classical energy 
\be
E_{\rm cl}=E_{\rm cl}[F]=\int d^3r \left\{\frac{f_\pi^2}{2}
\left(r^2\fpt+2\sFt\right)+\frac{\sFt}{2e^2}
\left(2\fpt+\frac{\sFt}{r^2}\right)\right\}\, .
\label{ecl}
\ee
It can be argued~\cite{Wi83}
that the baryon number equals the winding number associated with
the mapping~(\ref{hedgehog}), {\it i.e.} $B=[F(\infty)-F(0)]/\pi$. Hence
the boundary conditions $F(0)=-\pi$ and $F(\infty)=0$ corresponding
to unit baryon number determine the chiral angle uniquely. This soliton 
does not yet describe states of good spin and/or flavor as the 
{\it ansatz}~(\ref{hedgehog}) does not possess the corresponding 
symmetries. Such states are generated by restoring these symmetries 
through collective coordinates $A(t)$
\begin{equation}
U(\vec{r},t)\,=\,A(t)\, U_{\rm H}(\vec{r})\,A^\dagger(t)\, .
\label{collrot2}
\end{equation}
and subsequent canonical quantization thereof~\cite{Ad83}. This
introduces right $[A,R_i]=A\tau_i/2$ and left generators 
$[A,L_i]=\tau_i A/2$. While the isospin interpretation $I_i=L_i$ is 
general, the identity $J_i=-R_i$ for the spin is due to the hedgehog 
structure~(\ref{hedgehog}) as is the relation 
$|\vec{I}|=|\vec{J}|$. By quantizing the collective coordinates one
obtains a Hamiltonian in terms of physical observables
\begin{equation}
H_{\rm coll}\,=\,E_{\rm cl}+\frac{\vec{J}\,^2}{2\alpha^2}
\,=\,E_{\rm cl}+\frac{\vec{I}\,^2}{2\alpha^2}\, .
\label{hamrot2}
\end{equation}
The moment of inertia is also a functional of the above
determined chiral angle
\be
\alpha^2[F]=\frac{2}{3}\int d^3r\,
{\rm sin}^2F
\left[f_\pi^2+ \frac{1}{e^2}
\left(F^{\prime2}+\frac{{\rm sin}^2F}{r^2}\right)\right]\, .
\label{mominert}
\ee
Matching the empirical mass difference
$M_\Delta-M_{\rm N}=\frac{3}{2\alpha^2}
\sim 300{\rm MeV}$ fixes the so--far undetermined parameter to be
$e\approx4.0$.

\section{Extension to Three Flavors}

The generalization to three flavors is carried out straightforwardly
by taking $A(t)\in SU(3)$ with the hedgehog~(\ref{hedgehog}) embedded 
in the isospin subgroup. However, the Lagrangian becomes more
complicated as there are two essential extensions. The first one
is the Wess--Zumino--Witten term~\cite{Wi83}. Gauging it for local
$U_V(1)$ shows that indeed the winding number current equals the
baryonic current. Furthermore, as it is linear in the time derivative
it constrains $A$ to be quantized as a fermion (for $N_C$ odd). 
The second extension originates from flavor symmetry breaking that is
reflected by different masses and decay constants of the pseudoscalar
mesons
\begin{equation}
{\cal L}_{\rm SB}=\frac{f_\pi^2 m_\pi^2 -f_K^2 M_K^2}{2\sqrt3}
{\rm tr}\left\{\lambda_8\left(U+U^\dagger\right)\right\}
+\frac{f_K^2-f_\pi^2}{4\sqrt3}{\rm tr}\left\{\lambda_8
\left(\partial_\mu U \partial^\mu U^\dagger U 
+{\rm h.c.}\right)\right\}\, .
\label{lsymbr}
\end{equation}
The explicit form of ${\cal L}_{\rm SB}$ is model dependent, 
however, the techniques to study its effects on baryon properties are general.

In $SU(3)$ the collective coordinates are parameterized by 
\underline{eight} ``Euler--angles''
\be
A=D_2(\hat{I})\,{\rm e}^{-i\nu\lambda_4}D_2(\hat{R})\,
{\rm e}^{-i(\rho/\sqrt{3})\lambda_8}\ ,
\label{Apara}
\ee
where $D_2$ denote rotation matrices of three Euler--angles for each, 
rotations in isospace~($\hat{I}$) and coordinate--space~($\hat{R}$). 
Substituting the {\it ansatz}~(\ref{collrot2}) into 
${\cal L}+{\cal L}_{\rm SB}$ yields upon canonical quantization the 
Hamiltonian for the collective coordinates~$A$:
\be
H=H_{\rm s}+\fract{3}{4}\, \gamma\, {\rm sin}^2\nu\, .
\label{Hskyrme}
\ee
The symmetric piece of this collective Hamiltonian only contains
Casimir operators and may be expressed in terms of the $SU(3)$--right
generators $R_a\, (a=1,\ldots,8)$:
\be
H_{\rm s}=E_{\rm cl}+\frac{1}{2\alpha^2}\sum_{i=1}^3 R_i^2
+\frac{1}{2\beta^2}\sum_{\alpha=4}^7 R_\alpha^2\, .
\label{Hsym}
\ee
While $\beta^2$ is a moment of inertia 
similar to $\alpha^2$ in eq~(\ref{mominert}), $\gamma=\gamma[F]$ completely
originates from symmetry breaking
\bea
\gamma=
\frac{2\pi}{3}\int d^3r \left[\left(f_K^2m_K^2-f_\pi^2m_\pi^2\right)
\left(1-{\rm cos}F\right)+\frac{f_K^2-f_\pi^2}{2}
{\rm cos}F\left(F^{\prime2}r^2+2{\rm sin}^2F\right)\right].
\nonumber
\eea
The generators $R_a$ can be expressed in terms
of derivatives with respect to the `Euler--angles'. The eigenvalue
problem $H\Psi=\epsilon\Psi$ reduces to sets of ordinary second order
differential equations for isoscalar functions which only depend on
the strangeness changing angle $\nu$~\cite{Ya88} that can be integrated
numerically. Only the product
$\omega^2=\frac{3}{2}\gamma\beta^2$ appears in these differential
equations which is thus interpreted as the effective strength of the
flavor symmetry breaking. A value in the range 
$5{\scriptsize{\lapeq}}\omega^2{\scriptsize{\lapeq}}8$
is required to obtain reasonable agreement with the empirical mass
differences for the $\frac{1}{2}^+$ and $\frac{3}{2}^+$
baryons~\cite{We96}. The eigenstates of the symmetric piece~(\ref{Hsym})
are members of definite $SU(3)$ representations, {\it e.g.} the
octet ({\bf 8}) for the low--lying $\frac{1}{2}^+$ baryons. Upon flavor 
symmetry breaking, states of different representations are mixed.
At $\omega^2=6$ the nucleon amplitude contains a 23\% contamination
of the state with nucleon quantum numbers in the ${\bf \bar{10}}$ 
representation. That is, the baryon wave--functions exhibit 
strong distortion from flavor covariance. It is 
interesting to see what influence this has for baryon properties,
in particular the hyperon beta--decay for which flavor covariance
is known to work very well~\cite{Fl98}.

\section{Axial Current Matrix Elements}

\begin{table}[b]
~\vskip-0.5cm
\caption{\label{empirical}\sf The empirical values for the
$g_A/g_V$ ratios of hyperon beta--decays \protect\cite{DATA}.
For the process $\Sigma\to\Lambda$ only $g_A$ is given.
Also the flavor symmetric predictions are presented using the
values for $F$\&$D$ of Ref.~\protect\cite{Fl98}.}
~\vskip0.01cm
{\small
\begin{tabular}{ c || c | c | c | c | c}
& $\Lambda\to p$ & $\Sigma\to n$ & $\Xi\to\Lambda$ &
$\Xi\to\Sigma$ & $\Sigma\to\Lambda$\\
\hline
emp.& $0.718\pm0.015$ & $0.340\pm0.017$ & $0.25\pm0.05$ &
$1.287\pm0.158$ & $0.61\pm 0.02$\\
$F$\&$D$& $0.725\pm0.009$ & $0.339\pm0.026$ & $0.19\pm0.02$ &
$1.258=g_A$ & $0.65\pm0.01$
\end{tabular}}
\end{table}

The effect of the derivative type symmetry
breaking terms is mainly indirect. They provide the splitting between the
various decay constants and thus increase $\gamma$ because of
$f_K^2m_K^2-f_\pi^2m_\pi^2\approx 1.5f_\pi^2(m_K^2-m_\pi^2)$.
Otherwise the terms proprotional to $f_K^2-f_\pi^2$
may be omitted. Whence there are no symmetry
breaking terms in current operators and the non--singlet axial charge
operator is parameterized as 
\be
\int d^3r A_i^{(a)} = c_1 D_{ai} - c_2 D_{a8}R_i
+c_3\sum_{\alpha,\beta=4}^7d_{i\alpha\beta}D_{a\alpha}R_\beta
\, ,
\label{axsym}
\ee
where
$D_{ab}=\frac{1}{2}{\rm tr}\left(\lambda_a A\lambda_b A^\dagger\right)$,
$ a=1,\ldots,8$ and $i=1,2,3$.
In the limit $\omega^2\to\infty\, $
(integrating out {\it strange} degrees of freedom)
the strangeness contribution to the axial charge
of the nucleon should vanish. As the baryon wave--functions
parameterically depend on $\omega^2$ one finds
$\langle N| D_{83}| N\rangle\to0$ and
$\langle N| \sum_{\alpha,\beta=4}^7d_{3\alpha\beta}D_{8\alpha}R_\beta
| N\rangle\to0$ while $\langle N| D_{88}| N\rangle\to1$ for
$\omega^2\to\infty$. Thus the above consistency condition requires
\be
\int d^3r A_i^{(0)}= -2\sqrt{3} c_2 R_i\quad i=1,2,3\,.
\label{singsym}
\ee
for the singlet axial current because it yields a vanishing strangeness
projection, $A_i^{(s)}=(A_i^{(0)}-2\sqrt{3}A_i^{(8)})/3$ for 
$\omega^2\to\infty$. Note that the appearance of $c_2$ in 
eq~(\ref{singsym}) goes beyond group theoretical arguments.
Actually all model calculations in the literature \cite{Pa92,Bl93}
are consistent with (\ref{singsym}).
In order to completely describe the hyperon beta--decays we
also demand matrix elements of the vector charges. These are obtained
from the operator
\be
\int d^3r V_0^{(a)} = \sum_{b=1}^8D_{ab}R_b=L_a,
\label{vector}
\ee
which introduces the $SU(3)$--left generators $L_a$.

\begin{figure}[t]
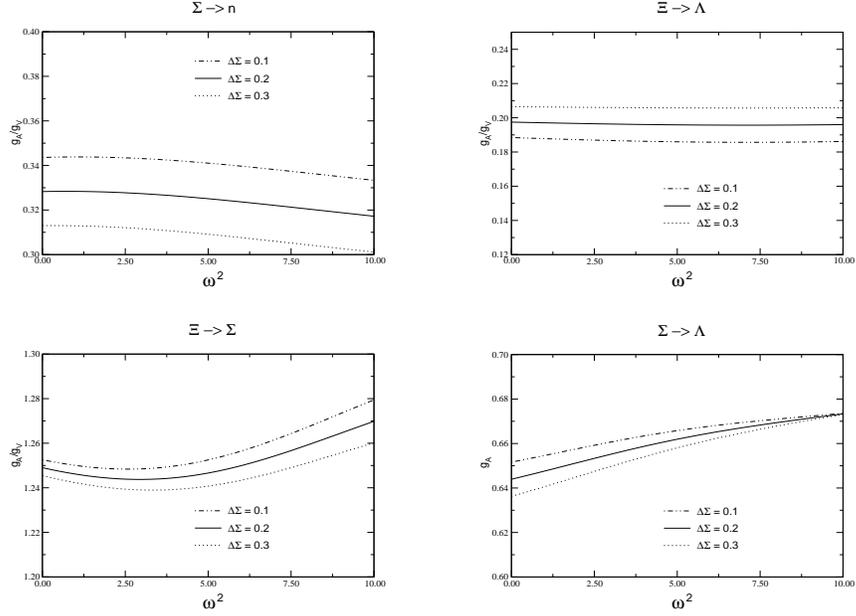

~\vskip0.05cm
\centerline{
\epsfig{figure=nusi.eps,height=5.0cm,width=3.8cm,angle=270}
\hspace{1.0cm}
\epsfig{figure=chla.eps,height=5.0cm,width=3.8cm,angle=270}}
~\vskip0.03cm
\centerline{
\epsfig{figure=sich.eps,height=5.0cm,width=3.8cm,angle=270}
\hspace{1.0cm}
\epsfig{figure=sila.eps,height=5.0cm,width=3.8cm,angle=270}}
\caption{\label{decay}\sf The predicted decay parameters for the
hyperon beta--decays using $\omega^2_{\rm fix}=6.0$.
The errors originating from those in $\Delta\Sigma_N$ are indicated.}
\end{figure}

The values for $g_A$ and $g_V$
(only $g_A$ for $\Sigma^+\to\Lambda e^+\nu_e$)
are obtained from the matrix elements of respectively
the operators in eqs~(\ref{axsym}) and~(\ref{vector}), sandwiched
between the eigenstates of the full Hamiltonian~(\ref{Hskyrme}).
I choose $c_2$ according to
$\langle N |\int d^3r A_3^{(0)} |N\rangle=-\sqrt3 c_2=
\Delta\Sigma=0.2\pm0.1$~\cite{El95}  and subsequently
determine $c_1$ and $c_3$ at $\omega^2_{\rm fix}=6.0$ such that the
emprical values for the nucleon axial charge, $g_A$ and the $g_A/g_V$ 
ratio for $\Lambda\to p e^-\bar{\nu}_e$ are reproduced\footnote{Here
the problem of the too small model prediction for $g_A$ will not be
addressed but rather the empirical value $g_A=1.258$ will be used
as an input to determine the $c_n$.}. This not only predicts the
other decay parameters but also describes the variation with
symmetry breaking as shown in figure~\ref{decay}.
The dependence on flavor symmetry breaking is very
moderate\footnote{However, the individual matrix elements
entering the ratios $g_A/g_V$ vary strongly with
$\omega^2$~\cite{We00}.} and the results can be viewed as reasonably
agreeing with the empirical data, {\it cf.} table \ref{empirical}.
The two transitions, $n\to p$ and $\Lambda\to p$, which are not shown in
figure~\ref{decay}, exhibit a similar neglegible dependence on $\omega^2$.
The observed independence of $\omega^2$ shows that these predictions
are not sensitive to the choice of $\omega^2_{\rm fix}$.
Comparing the results in figure \ref{decay} with the data in
table~\ref{empirical} shows that the calculation using the strongly
distorted wave--functions agrees equally well with the empirical
data as the flavor symmetric $F$\&$D$ fit. 

\begin{figure}[t]
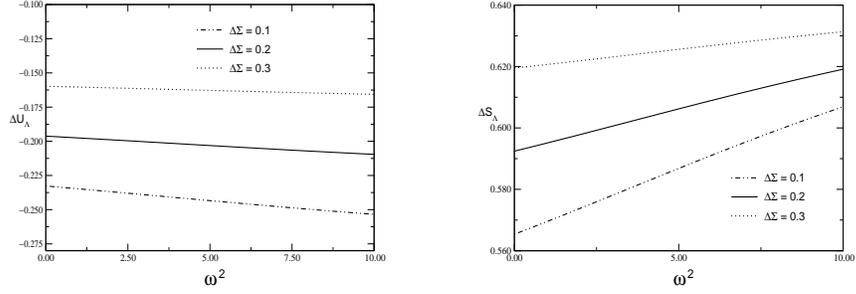

~\vskip-0.03cm
\centerline{
\epsfig{figure=h1la.eps,height=5.0cm,width=3.8cm,angle=270}
\hspace{1.0cm}
\epsfig{figure=h3la.eps,height=5.0cm,width=3.8cm,angle=270}}
\caption{\label{laxial}\sf The contributions of the {\it non--strange}
(left panel) and {\it strange} (right panel) degrees of freedom
to the axial charge of the $\Lambda$. Again $\omega^2_{\rm fix}=6.0$ 
was assumed.}
\end{figure}

Figure \ref{laxial} shows the flavor components of the axial
charge of the $\Lambda$ hyperon. Again, the various contributions
to the axial charge of the $\Lambda$ exhibit only a moderate dependence
on $\omega^2$. The {\it non--strange} component,
$\Delta U_\Lambda=\Delta D_\Lambda$ slightly increases in magnitude.
The {\it strange} quark piece, $\Delta S_\Lambda$ grows with
symmetry breaking since $\Delta\Sigma_\Lambda$ is kept fixed. These
results agree nicely with an $SU(3)$ analysis applied to the
data~\cite{Ja96a}. The observed independence on $\omega^2$ does not 
occur for all matrix elements of the axial current. A prominent exemption
is the {\it strange} quark component in the nucleon, $\Delta S_N$.
For $\Delta\Sigma=0.2$, say, it is significant at zero symmetry
breaking, $\Delta S_N=-0.131$ while it decreases (in magnitude) to
$\Delta S_N=-0.085$ at $\omega^2=6.0$.

This far I have only considered the general sturcture of the
current operators in soliton models without actually computing
the constants $c_i$ from a model soliton, although I had the Skyrme model 
in mind. However, this model is too simple to be realistic. For 
example, it improperly predicts $\Delta\Sigma=0$~\cite{We96} and
more complicted models must be utilized, as {\it e.g.}
the vector meson model that has been established for 
two flavors in ref~\cite{Ja88}. Later it has been generalized to 
three flavors and been shown to fairly describe hyperon
beta--decay~\cite{Pa92}. A minimal set of
symmetry breaking terms is included \cite{Ja89} to account
for different masses and decay constants. These terms add
symmetry breaking pieces to the axial charge operator,
$$
\delta A_i^{(a)}=c_4 D_{a8}D_{8i}+
c_5\hspace{-1.0mm} \sum_{\alpha,\beta=4}^7
d_{i\alpha\beta}D_{a\alpha}D_{8\beta}+
c_6 D_{ai}(D_{88}-1)\,\, ,\,
\delta A_i^{(0)}= 2\sqrt{3}\,c_4D_{8i}\, .
$$
The coefficients $c_1,\ldots,c_6$ are functionals of the soliton 
and can be computed once the soliton is constructed~\cite{We00}.
As the model parameters cannot be completely determined in the 
meson sector~\cite{Ja88} I use the small remaining freedom to 
accommodate baryon properties in three different ways, see
table \ref{realistic}. The set denoted by `masses' refers 
to a best fit to the baryon mass differences.
It predicts the axial charge somewhat on the low side, $g_A=0.88$.
The set named `mag.mom.' refers to parameters that yield 
magnetic moments of the $\frac{1}{2}^+$ baryons
close to the respective empirical data
(with $g_A=0.98$) and finally the set labeled `$g_A$' reproduces the
axial charge of the nucleon and also reasonably accounts for hyperon
beta--decay~\cite{Pa92}.
\begin{table}[t]
\caption{\label{realistic}\sf Spin content of the $\Lambda$ in the
realistic vector meson model. For comparison the nucleon
results are also given. Three sets of model parameters
are considered, see text.}
~\vskip0.01cm
{\small
\begin{tabular}{ c || c | c |c || c | c | c | c}
& \multicolumn{3}{c||}{$\Lambda$} &
\multicolumn{4}{c}{$N$}\\
\hline
& $\Delta U = \Delta D$ & $\Delta S$ & $\Delta\Sigma$ &
 $\Delta U$ & $\Delta D$ & $\Delta S$ & $\Delta\Sigma$\\
\hline
masses
&$-0.155$&$0.567$&$0.256$&$0.603$&$-0.279$&$-0.034$&$0.291$\\
mag. mom.
&$-0.166$&$0.570$&$0.238$&$0.636$&$-0.341$&$-0.030$&$0.265$\\
$g_A$
&$-0.164$&$0.562$&$0.233$&$0.748$&$-0.476$&$-0.016$&$0.256$
\end{tabular}}
\end{table}
As presented in table~\ref{realistic}, the predictions for the axial 
properties of the $\Lambda$ are insensitive to the model parameters.
The singlet matrix element of the $\Lambda$ hyperon is 
smaller than that of the nucleon. Sizable polarizations of the {\it up} 
and {\it down} quarks in the $\Lambda$ are predicted. They are slightly 
smaller in magnitude but nevertheless comparable to those obtained 
from the $SU(3)$ symmetric analyses~\cite{Ja96a}.

\section{Solitons in a Chiral Quark Model}

In the preceding discussions I have illuminated the success of the
soliton picture using hyperon--beta decay and axial current matrix
elements as examples. To study those processes it was sufficient 
to consider static baryon properties as computed from a purely 
meson model Lagrangian. In order to advance and compute structure 
functions as they are measured in deep inelastic scattering (DIS) in
the soliton picture it will be necessary to trace the quark degrees
of freedom as the meson model is constructed. As already mentioned,
this cannot be accomplished in QCD and therefore I will utilize
a simpler model of the quark flavor dynamics. 
Integrating out the quark degrees gives a bosonized action
that can be cast in the form
\be
{\cal A} [S,P]=-iN_C{\rm Tr}_{\textstyle\Lambda}{\rm log}\, 
\left[i\dslash-\left(S+i\gamma_5P\right)\right]
-\frac{1}{4G}\int d^4x\, {\rm tr}\, {\cal V}(S,P)\, .
\label{bosact}
\ee
Here ${\cal V}$ is a local potential respectively for scalar and 
pseudoscalar fields $S$ and $P$ which are matrices in flavor space.
For example, in the Nambu--Jona--Lasinio (NJL) model~\cite{Na61} one
has ${\cal V}=S^2+P^2+2{\hat m}_0S$ with $\hat{m}_0$ being the
current quark mass matrix. The functional trace 
(${\rm Tr}$) refects integrating out the quarks and induces a non--local
interaction for $S$ and $P$. A major concern in regularizing the 
functional (\ref{bosact}), as indicated by the cut--off $\Lambda$, is to 
maintain the chiral anomaly. This is achieved by splitting this functional 
into $\gamma_5$--even and odd pieces via
\begin{eqnarray}
&&{\rm Tr}_\Lambda {\rm log}\,
\left[i\dslash-\left(S+i\gamma_5P\right)\right]
=-i\frac{N_C}{2} \sum_{i=0}^2 c_i {\rm Tr}\, {\rm log}
\left[- \bD \bD_5 +\Lambda_i^2-i\epsilon\right]
\hspace{1.27cm} \nonumber \\ &&\hspace{5cm}
-i\frac{N_C}{2}
{\rm Tr}\, {\rm log}
\left[-\bD \left(\bD_5\right)^{-1}-i\epsilon\right]\, ,
\hspace{1cm}
\label{PVreg} \\
{\rm with}&&\qquad
i \bD = i\dslash - \left(S+i\gamma_5P\right) 
\quad {\rm and} \quad
i \bD_5 = - i\dslash - \left(S-i\gamma_5P\right)\, .
\label{defd}
\end{eqnarray}
With the conditions
$ c_0=1\, ,\,\, \Lambda_0=0 ,\, \sum_{i=0}^2c_i=0\,
{\rm and}\, \sum_{i=0}^2c_i\Lambda_i^2=0$ the double
Pauli--Villars regularization renders the functional
(\ref{bosact}) finite. The $\gamma_5$--odd piece is
conditionally finite and not regularizing it properly
reproduces the chiral anomaly.  For sufficiently large $G$ one 
obtains the VEV, $\langle S\rangle\equiv m\ne0$ that parameterizes 
the dynamical chiral symmetry breaking from the gap--equation. 

Before discussing nucleons as solitons of the bosonized 
action~(\ref{bosact}) and the respective structure functions
it will be illuminating to first consider DIS off pions.
Details of these studies are published in \cite{We99}. 
For related work see refs~\cite{We96a,Di96,Wa98}.

\section{The Compton Tensor and Pion Structure Function}
DIS off hadrons is parameterized by 
the hadronic tensor $W^{\mu\nu}(q)$ with $q$ being the momentum 
transmitted to the hadron by the photon. $W^{\mu\nu}(q)$ is obtained from the 
hadron matrix element of the commutator $[J^\mu(\xi), J^\nu(0)]$. 
An essential feature of bosonized quark models is that
the derivative term in (\ref{bosact}) is formally identical to that of
a non--interacting quark model. Hence the current operator is given as
$J^\mu={\bar q}{\cal Q}\gamma^\mu q$, with ${\cal Q}$ being a flavor
matrix. Expectation values of currents are computed by
introducing pertinent sources $v_\mu$ in eq~(\ref{PVreg}) 
\be
i\bD\, \longrightarrow \, i\bD +{\cal Q}\vslash
\qquad {\rm and} \qquad
i\bD_5\, \longrightarrow \, i\bD_5 -{\cal Q}\vslash
\label{source}
\ee
and taking appropriate derivatives.
In bosonized quark models it is convenient to start from the 
absorptive part of the forward virtual Compton 
amplitude\footnote{The momentum of the hadron is called $p$ and 
its spin eventually $s$.} 
\be
T^{\mu\nu}(q)=\int d^4\xi\, {\rm e}^{iq\cdot\xi}\,
\langle p,s|T\left(J^\mu(\xi) J^\nu(0)\right)|p,s\rangle
\quad {\rm and }\quad
W^{\mu\nu}(q)=\frac{1}{2\pi} \Im\, (T^{\mu\nu}(q)\, .
\label{comp1}
\ee
because the time--ordered product is 
unambiguously obtained from 
\be
T\left(J^\mu(\xi) J^\nu(0)\right)=
\frac{\delta^2}{\delta v_\mu(\xi)\, \delta v_\nu(0)}\,
{\rm Tr}_\Lambda {\rm log}\,
\left[i\dslash-\left(S+i\gamma_5P\right)+{\cal Q}\,
\vslash\right]\Big|_{v_\mu=0}\, \, ,
\label{tprod}
\ee
as defined from eqs~(\ref{PVreg}) with the substitution~(\ref{source}).
In order to extract the leading twist pieces of the structure 
functions, $W^{\mu\nu}(q)$ is studied in the Bjorken limit: 
$q^2\to-\infty$ with $x=-q^2/p\cdot q$ fixed.

DIS off pions is characterized by a single structure function,
$F(x)$. For its computation the pion matrix element in the Compton 
amplitude~(\ref{comp1}) must be specified. Whence I 
introduce the pion field ${\vec\pi}$ via\footnote{The coupling $g$ 
and the constituent quark mass $m$ are related by the pion decay 
constant. In the chiral limit the relation is linear: $m=g f_\pi$.}
\be
S+iP\gamma_5=m\, \left(U\right)^{\gamma_5} = m\, {\rm exp}
\left(i \frac{g}{m}\gamma_5\,{\vec\pi} \cdot {\vec\tau} \right)\, .
\label{SandP}
\ee
In a first step pion properties are utilized to fix the 
model parameters. It is also worthwhile to mention that
expanding (\ref{PVreg},\ref{SandP}) to linear and quadratic order  
in $\vec{\pi}$ and $v_\mu$, respectively yields the proper result for
the anomalous decay $\pi^0\to\gamma\gamma$. The Compton amplitude 
for virtual pion--photon scattering is obtained 
by expanding (\ref{PVreg},\ref{SandP}) to second order in both, 
${\vec\pi}$ and $v_\mu$. Due to the separation into $\bD$ and $\bD_5$ 
this calculation differs considerably from the evaluation of the 
`handbag' diagram because isospin violating dimension--five operators 
emerge. Fortunately all isospin violating pieces cancel yielding 
$$
F(x)=\frac{5}{9} (4N_C g^2)
\frac{d}{dm_\pi^2}\left\{m_\pi^2
\sum_{i=0}^2 c_i\, \frac{d^4k}{(2\pi)^4i}\,
\left[k^2+x(1-x)m_\pi^2-m^2-\Lambda_i^2+i\epsilon\right]^{-2}\right\}\, .
\nonumber
$$
The cancellation of the isospin violating pieces is a feature
of the Bjorken limit: insertions of the pion field on the propagator
carrying the infinitely large photon momentum can be safely 
ignored. Furthermore this propagator can be taken to be the one 
for non--interacting massless fermions. This implies that also
the Pauli--Villars cut--offs can be omitted for this propagator
leading to the desired scaling behavior of the structure function.

\section{Nucleon Structure Functions}
Assuming the hedgehog shape~(\ref{hedgehog}) for the pion
field~(\ref{SandP}) the profile function $F(r)$ enters the single 
particle Dirac Hamiltonian
\be
h=\vec{\alpha}\cdot\vec{p} +\beta\, m\, \left[{\rm cos}F+
i\gamma_5 \vec{\tau}\cdot\hat{r}\, {\rm sin}F\right]\, .
\label{Dirac}
\ee
Denoting its eigenvalues with $\epsilon_\alpha$ allows one to
construct an energy functional~\cite{Do92}
\bea
E[F]&=&
\frac{N_C}{2}\left(1-{\rm sign}(\epsilon_{\rm val})\right)
\epsilon_{\rm val}
-\frac{N_C}{2}\sum_{i=0}^2 c_i \sum_\alpha
\left\{\sqrt{\epsilon_\alpha^2+\Lambda_i^2}
-\sqrt{\epsilon_\alpha^{(0)2}+\Lambda_i^2}
\right\}
\nonumber \\ && \hspace{2cm}
+m_\pi^2f_\pi^2\int d^3r \, (1-{\rm cos}F)\, 
\label{etot}
\eea
for a unit baryon number configuration.
Here $\epsilon_{\rm val}$ denotes the unique bound state level and
$\epsilon_\alpha^{(0)}$ are the eigenvalues when the soliton
is absent. The soliton profile $F(r)$ is then obtained from extremizing 
$E$ self--consistently~\cite{Al96}. As described in sections~2 and~3, 
nucleon states are generated by introducing collective 
coordinates~(\ref{collrot2}) and subsequent canonical quantization
thereof~\cite{Re89}. 

As argued in the previous section the quark propagator with the infinite 
photon momentum should be taken to be free and massless. Thus, it is 
sufficient to differentiate (Here $\bD$ and $\bD_5$ are those
of eq~(\ref{defd}), {\it i.e.} with $v_\mu=0$.)
\begin{eqnarray}
&&
\hspace{-0.6cm}
\frac{N_C}{4i}\sum_{i=0}^2c_i
{\rm Tr}\,\left\{\left(-\bD\bD_5+\Lambda_i^2\right)^{-1}
\left[{\cal Q}^2\vslash\left(\dslash\right)^{-1}\vslash\bD_5
-\bD(\vslash\left(\dslash\right)^{-1}\vslash)_5
{\cal Q}^2\right]\right\}
\nonumber \\ &&
+\frac{N_C}{4i}
{\rm Tr}\,\left\{\left(-\bD\bD_5\right)^{-1}
\left[{\cal Q}^2\vslash\left(\dslash\right)^{-1}\vslash\bD_5
+\bD(\vslash\left(\dslash\right)^{-1}\vslash)_5
{\cal Q}^2\right]\right\}\, ,\,\,
\label{simple}
\end{eqnarray}
with respect to the photon field $v_\mu$. 
I have introduced the $(\ldots)_5$ description
\be
\gamma_\mu\gamma_\rho\gamma_\nu
=S_{\mu\rho\nu\sigma}\gamma^\sigma
-i\epsilon_{\mu\rho\nu\sigma}\gamma^\sigma\gamma^5
\quad {\rm and} \quad
(\gamma_\mu\gamma_\rho\gamma_\nu)_5
=S_{\mu\rho\nu\sigma}\gamma^\sigma+
i\epsilon_{\mu\rho\nu\sigma}\gamma^\sigma\gamma^5
\label{defsign}
\ee
to account for the unconventional appearance of axial 
sources in $\bD_5$~\cite{We99}. Substituting (\ref{collrot2})
into (\ref{simple}) and computing the functional trace, using a basis
of quark states obtained from the Dirac Hamiltonian~(\ref{Dirac}),
yields analytical results for the structure 
functions. I refer to \cite{We99} for detailed formulae and the 
explicit verification of sum rules. Here I simply report the important 
result that the structure function entering the Gottfried sum rule 
is related to the $\gamma_5$--odd piece of the action and hence does not 
undergo regularization. This is surprising because in the parton model 
this structure function differs from the one associated with the Adler 
sum rule only by the sign of the anti--quark distribution. The latter 
structure function, however, gets regularized, in agreement with the
quantization rules for the collective coordinates. 

\begin{figure}[t]
\begin{center}
\epsfig{figure=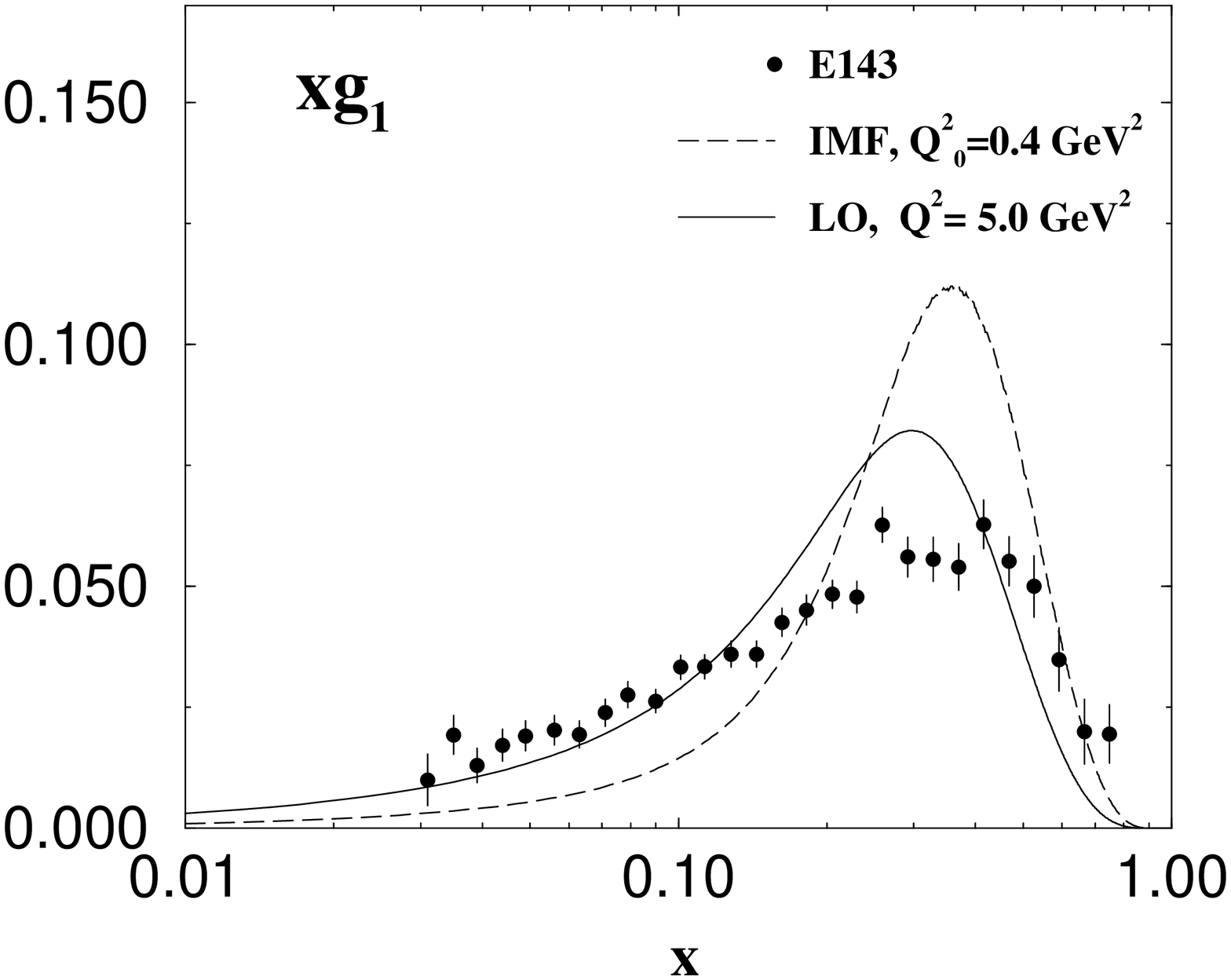,height=5.5cm,width=4.3cm,angle=270}
\hspace{1cm}
\epsfig{figure=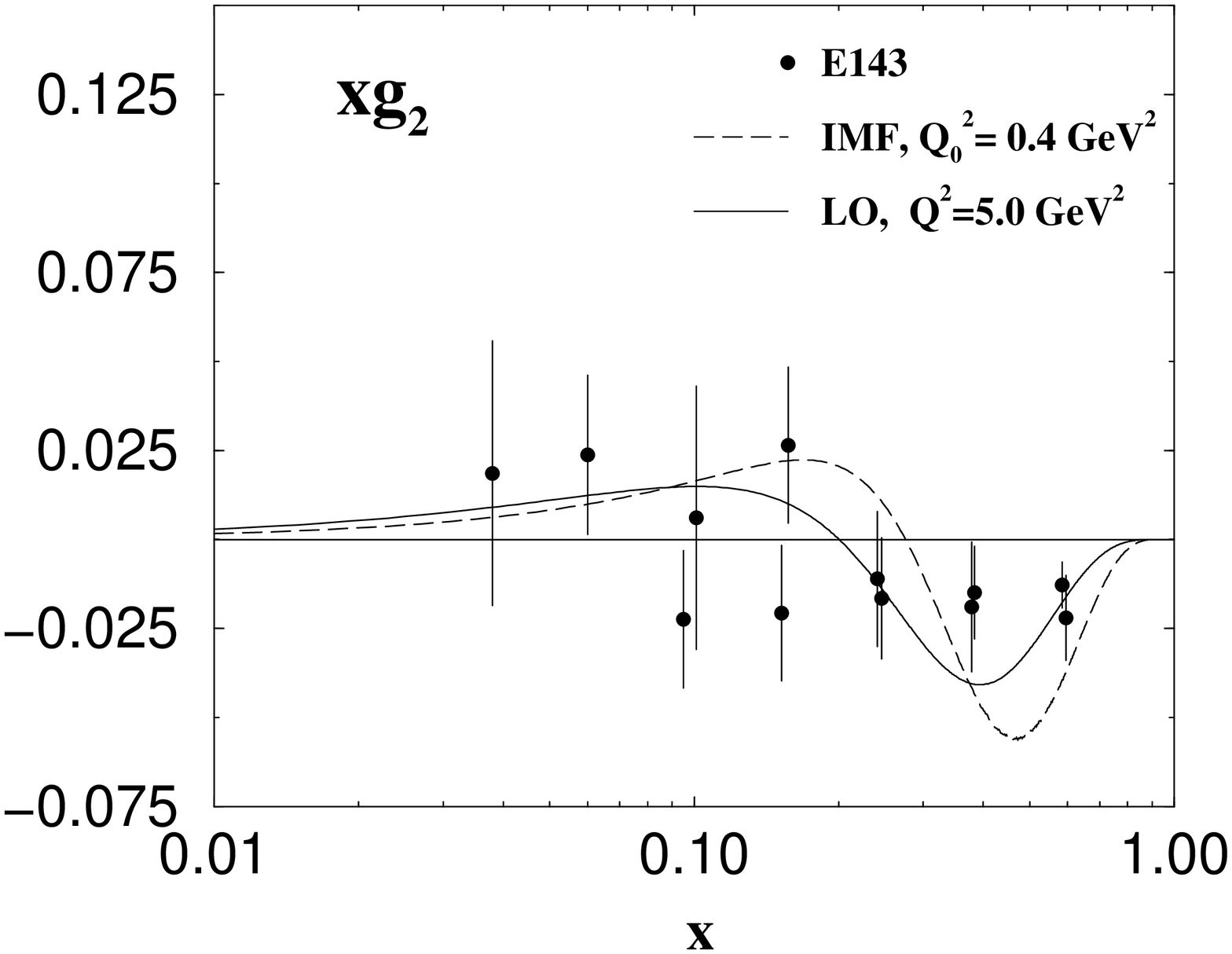,height=5.5cm,width=4.3cm,angle=270}
\end{center}
\vskip-0.5cm
\caption{\label{xg12}{\sf Model predictions for the
polarized proton structure functions $xg_1$ (left panel)
and $xg_2$ (right panel). The curves labeled `RF' denote the
results as obtained from the valence quark contribution to
(\protect\ref{simple}). These undergo a projection to the infinite
momentum frame `IMF' \cite{Ga98} and a leading order `LO'
DGLAP evolution \cite{DGLAP}. 
Data are from SLAC--E143 \cite{Abe98}. }}
\end{figure}

Unfortunately numerical results for the full structure functions 
in the double Pauli--Villars regularization scheme,
{\it i.e.} including the properly regularized vacuum piece are not yet 
available. However, in the Pauli--Villars regularization the axial 
charges are saturated to 95\% or more by their valence quark 
($\epsilon_{\rm val}$) contributions once the self--consistent soliton 
is substituted. This provides sufficient justification to adopt
the valence quark contribution to the polarized structure functions
as a reliable approximation \cite{We96a}. Note that the zeroth moment
of the leading structure function~$g_1$ is nothing but the axial current
matrix elements discussed in section~4. In figure~\ref{xg12} I compare 
the model predictions for the linearly independent polarized structure 
functions to experimental data~\cite{Abe98}.
The evolution of the structure function $g_2$ to the momentum scale of 
the experiments requires the separation into twist--2 and twist--3 
components \cite{DGLAP}. The model results for the 
polarized structure functions, which I argued to have reliably approximated, 
agree reasonably well with the experimental data. 

\section{Conclusions}

In these lectures I have discussed a twofold program to
study the nucleon spin structure in chiral soliton models.
After having reviewed the arguments from large $N_C$ QCD
that lead to the picture that baryons emerge as solitons
in an effective meson theory I have adopted that picture
to compute various baryon matrix elements. Here I focused 
on the effects of flavor symmetry breaking in the baryon
wave--functions and showed that despite of strong deviations
from flavor covariant wave--functions the empirical parameters
for hyperon beta--decay are reproduced. I also showed that
chiral soliton models provide a consistent explanation of
the proton spin puzzle, {\it i.e.} the smallness of the observed
axial singlet current matrix element. On the other hand
flavor symmetry breaking in the nucleon wave--function leads
to a significant reduction of the polarization of the strange
quarks inside the nucleon. In the second part of these lectures
I describe how an effective meson theory that contains chiral
soliton solutions can be constructed from a simplified model for 
the quark flavor dynamics. Since the quark degrees of freedom can
be traced through this bosonization procedure it is possible to 
compute nucleon structure functions from this soliton. It turns 
out that additional correlations are introduced due to the 
unavoidable regularization which is imposed in a way to respect the
chiral anomaly. Hence a consistent extraction of the nucleon structure 
functions from the Compton amplitude in the Bjorken limit leads to 
expressions that are quite different from those obtained by an 
{\it ad hoc} regularization of quark distributions in the same 
model. I also showed that within a reliable 
approximation the numerical results for the spin dependent 
structure functions agree reasonably well with the empirical 
data.

\bigskip
{\small I would like to thank Miroslav Finger for organizing
this worthwhile Advanced Study Institute. 
This work has been supported by the Deutsche Forschungsgemeinschaft
under contracts We 1254/3-1 and We 1254/4-2.}

\end{document}